\documentclass[conference]{IEEEtran}
\IEEEoverridecommandlockouts
\usepackage{cite}
\usepackage{amsmath,amssymb,amsfonts}
\usepackage{algorithmic}
\usepackage{graphicx}
\usepackage{textcomp}
\usepackage{xcolor}
\def\BibTeX{{\rm B\kern-.05em{\sc i\kern-.025em b}\kern-.08em
    T\kern-.1667em\lower.7ex\hbox{E}\kern-.125emX}}
\begin{document}

\title{Performance Evaluation of eLoran Spatial ASF Corrections Based on Measured ASF Map
\thanks{This work was supported in part by Grant RS-2024-00407003 from the ``Development of Advanced Technology for Terrestrial Radionavigation System'' project, funded by the Ministry of Oceans and Fisheries, Republic of Korea;
in part by the National Research Foundation of Korea (NRF), funded by the Korean government (Ministry of Science and ICT), under Grant RS-2024-00358298; 
in part by the Future Space Navigation and Satellite Research Center through the NRF, funded by the Ministry of Science and ICT (MSIT), Republic of Korea, under Grant 2022M1A3C2074404; 
and in part by the MSIT, Korea, under the Information Technology Research Center (ITRC) support program supervised by the Institute of Information \& Communications Technology Planning \& Evaluation (IITP) under Grant IITP-2024-RS-2024-00437494.
}
}

\author{\IEEEauthorblockN{Jaewon Yu} 
\IEEEauthorblockA{\textit{School of Integrated Technology} \\
\textit{Yonsei University}\\
Incheon, Korea \\
jaewon.yu@yonsei.ac.kr} 
\and
\IEEEauthorblockN{Pyo-Woong Son${}^{*}$} 
\IEEEauthorblockA{\textit{Department of Electronics Engineering} \\
\textit{Chungbuk National University} \\
Cheongju, Korea \\
pwson@cbnu.ac.kr}
{\small${}^{*}$ Corresponding author}
}

\maketitle

\begin{abstract}
This paper analyzes the effectiveness of spatial ASF correction methods in the Korean eLoran system using measured ASF maps. 
Three correction scenarios were evaluated under identical simulation settings: no correction (S0), local correction using true ASF values (S1), and wide-area correction using a single reference station value (S2). 
Simulation results show that S1 consistently achieved the lowest positioning errors, while S0 exhibited the largest errors with extensive high-error regions. 
S2 provided limited improvements near the reference station but degraded with increasing distance and ASF spatial gradients. 
The findings highlight that local ASF correction significantly improves eLoran positioning performance, whereas wide-area correction has only localized benefits.
\end{abstract}


\section{Introduction}

Global Navigation Satellite Systems (GNSS) \cite{Kim14:Comprehensive, Chen11:Real, Lee23:Seamless, Kim23:Machine, Kim23:Low, Lee24:A, Kim23:Single, Kim22:Machine, Lee22:Urban, Kim25:Set, Jeong24:Quantum} are essential infrastructures that provide worldwide Positioning, Navigation, and Timing (PNT) services, supporting aviation, maritime, and telecommunication sectors. 
However, due to their inherently low transmission power, GNSS signals have long been recognized as vulnerable to intentional interference such as jamming and spoofing\cite{Park21:Single, Park18:Dual, Kim19:Mitigation, Park17:Adaptive, Jeong20:RSS, Moon24:HELPS, Lee22:Performance, Park25:Toward}.  

According to a C4ADS analysis, between February 2016 and November 2018, a total of 9,883 spoofing incidents were confirmed worldwide, affecting 1,311 merchant vessels \cite{C4ADS19:Above}. 
Since 2022, cases of GNSS jamming and spoofing have sharply increased on a global scale, with official reports issued by the UK Civil Aviation Authority (CAA) \cite{CAA24:Global}. 
Furthermore, since 2010, strong interference signals transmitted from North Korea have repeatedly disrupted aviation, maritime, and telecommunication services across Incheon and the West Sea, confirming the severe impact of GPS jamming on critical national infrastructure \cite{Kim22:First, Rhee21:Enhanced, Son20:eLoran, Son24:eLoran}.  

As a result, the eLoran system \cite{Di2025:longwave, Yin25:Preliminary, Yang25:Research, Cheng25:Research}, a terrestrial navigation system with high transmission power, has gained attention as a potential backup to GPS, and research on eLoran has been actively pursued in Korea \cite{Son23:Demonstration, Son22:Compensation, Kang25:Enhancing}. 
During signal propagation, eLoran is subject to three types of delays: Primary Factor (PF), Secondary Factor (SF), and Additional Secondary Factor (ASF). 
PF and SF can be theoretically approximated using Brunavs’ equations, but ASF is strongly influenced by terrain and ground conductivity, making accurate prediction through simple calculation highly challenging\cite{Di25:Comparative, Hehenkamp2023:Prediction}.  

To address this, ASF is divided into temporal ASF, which varies over time, and spatial ASF, which remains constant over time. 
Temporal ASF can be compensated in real-time by reference stations, while spatial ASF is corrected using ASF maps based on pre-surveyed data. 
Since temporal ASF generally exhibits smaller variation, accurate estimation of spatial ASF plays an important role in improving overall correction performance.  

In practice, spatial ASF is typically corrected by interpolating pre-surveyed data to generate ASF maps, or by broadcasting single correction values derived at reference stations over a wide area \cite{Son19:Universal, Williams00:ASF,Hwang18:TDOA}. 
However, spatial ASF varies significantly depending on land-to-sea ratio, terrain and altitude changes, and effective ground conductivity. 
Applying a single correction value across a wide region inevitably produces residual errors, which accumulate with distance and terrain variation. 
Therefore, it is necessary to evaluate, within a simulation environment, the errors associated with applying a single correction value compared to those obtained using ASF maps, in order to clarify both the limitations and potential improvements of spatial ASF correction methods.  

This paper assumes the measured ASF map as the ground truth and compares three scenarios under identical conditions. 
Scenario S0 (no correction) leaves spatial ASF uncorrected, allowing path-dependent offsets to propagate directly into positioning errors. 
Scenario S1 (local ASF correction) removes the true ASF value at each grid point, resulting in near-zero residual errors as an idealized case. 
Scenario S2 (single-value correction) applies the ASF value from a reference station uniformly across the entire area, with residuals defined as the difference between the true ASF and the reference ASF. 
In S2, residual errors increase with the spatial gradient of ASF and distance from the reference station, leading to acceptable performance only near the station but greater degradation in regions with steep gradients or greater distances.  

By visualizing the positioning error distributions for these three scenarios and quantitatively comparing their performance, this study provides practical evidence regarding the effectiveness and limitations of spatial ASF correction methods.

\section{Methodology}

\subsection{Simulation Setup}

In this study, the LCAST simulator was used to compute grid-based positioning accuracy under each correction scenario. 
The spatial ASF map \cite{NMPO23:ASF} used in this study was produced by National Maritime PNT Office in Korea and is based on measured values interpolated across the study area. This map is assumed as the ground truth for all analyses.  
All three scenarios were evaluated under identical noise, jitter, SNR threshold, transmitter configuration, and grid resolution to ensure that only the correction method itself contributed to performance differences. 
Specifically, the simulator grid coordinates were mapped onto the ASF map , and the ASF$_{\text{true}}$ values at each simulator grid point were obtained through linear interpolation. 
This initialization was kept identical across all scenarios. 
The simulation parameters were configured as summarized in Table~\ref{tab:parameters}.

\begin{table}
\centering
\caption{Simulation Input Parameters}
\begin{tabular}{|l|c|}
\hline
\textbf{Simulation Input Parameters} & \textbf{Settings} \\
\hline
Pohang Transmitter Power      & 150kW \\
Gwangju Transmitter Power         & 50kW \\
Socheong Transmitter Power       & 8kW \\
Pohang Transmitter Jitter     & 2.11m \\
Gwangju Transmitter Jitter        & 3.21m \\
Socheong Transmitter Jitter      & 2.11m \\
Season                          & 'Averaged' \\
Noise Level                     & 95\% \\
SNR Threshold                   & -15dB \\
\hline
\end{tabular}
\label{tab:parameters}
\end{table}

\subsection{Scenario Definition}

Three correction scenarios were compared under identical transmission and reception conditions.  

\textbf{S0 (No Correction)} serves as the baseline, in which spatial ASF is not removed and path-dependent ASF offsets directly propagate into positioning errors.  

\textbf{S1 (Local Correction)} represents an ideal case, where the true spatial ASF value of each grid point is perfectly removed, resulting in residuals of approximately $\delta \mathrm{ASF} \approx 0$.  

\textbf{S2 (Wide-Area Correction)} assumes an operational case where a single ASF value at the reference station is uniformly applied across the entire area. The path-dependent residual is defined as
\begin{equation}
\delta \mathrm{ASF}(\text{lat}, \text{lon}, \text{stn}) 
= \mathrm{ASF}_{\text{true}}(\text{lat}, \text{lon}, \text{stn}) 
- \mathrm{ASF}_{\text{ref}}(\text{stn}),
\end{equation}
where $\mathrm{ASF}_{\text{true}}$ denotes the value from the measured ASF map, and 
\begin{equation}
\mathrm{ASF}_{\text{ref}}(\text{stn}) 
= \mathrm{ASF}_{\text{true}}(\text{lat}_{\text{ref}}, \text{lon}_{\text{ref}}, \text{stn})
\end{equation}
is the ASF value extracted from the map at the Incheon reference station.  

\subsection{Accuracy Computation Based on ASF Residuals}

Positioning accuracy (ACC) at each grid point was evaluated using three transmitters: Pohang, Gwangju, and Socheong. Scenario-dependent ASF residuals are given by

\begin{itemize}
  \item \textbf{S0:} $r_s = \mathrm{ASF}_{\text{true}}(s)$
  \item \textbf{S1:} $r_s = 0$
  \item \textbf{S2:} $r_s = \mathrm{ASF}_{\text{true}}(s) - \mathrm{ASF}_{\text{ref}}(s)$
\end{itemize}

and collected into a residual vector
\[
\mathbf{r}_{\text{ASF}} = [r_1, r_2, r_3]^\top.
\]

These residuals are converted into range biases by multiplying the speed of light $c$:
\[
\mathbf{d} = c \cdot \mathbf{r}_{\text{ASF}}.
\]

The weighted least-squares formulation is then constructed with the geometry matrix $G$ and weight matrix $W$:
\[
M = G^\top W G.
\]

The random error component of the position estimate is given by
\[
\sigma_{\text{pos}} = \sqrt{(M^{-1})_{11} + (M^{-1})_{22}},
\]
while the ASF-induced position bias is obtained as
\[
\begin{bmatrix}\Delta x \\ \Delta y \\ \Delta b \end{bmatrix}
= M^{-1} G^\top W \mathbf{d}, 
\qquad
\text{pos\_bias} = \sqrt{(\Delta x)^2 + (\Delta y)^2}.
\]

Finally, the overall ACC is defined as
\[
\mathrm{ACC} = \sqrt{\sigma_{\text{pos}}^2 + \text{pos\_bias}^2}.
\]

Thus:
\begin{itemize}
\item \textbf{S0}: ASF residuals remain as $\mathbf{r}_{\text{ASF}}=\mathrm{ASF}_{\text{true}}$, yielding the largest $\text{pos\_bias}$ and the worst ACC.
\item \textbf{S1}: ASF residuals are eliminated ($\mathbf{r}_{\text{ASF}}=\mathbf{0}$), resulting in $\text{pos\_bias}=0$ and ACC determined only by $\sigma_{\text{pos}}$, the ideal lower bound.
\item \textbf{S2}: ASF residuals are $\mathbf{r}_{\text{ASF}}=\mathrm{ASF}_{\text{true}}-\mathrm{ASF}_{\text{ref}}$, so $\text{pos\_bias}$ grows with the spatial gradient $\nabla \mathrm{ASF}$ and the distance from the reference station, leading to gradual degradation of ACC.
\end{itemize}

\section{Results}
\subsection{Comparison of ACC Maps}
\begin{figure}
    \centering
    \includegraphics[width=0.8\linewidth]{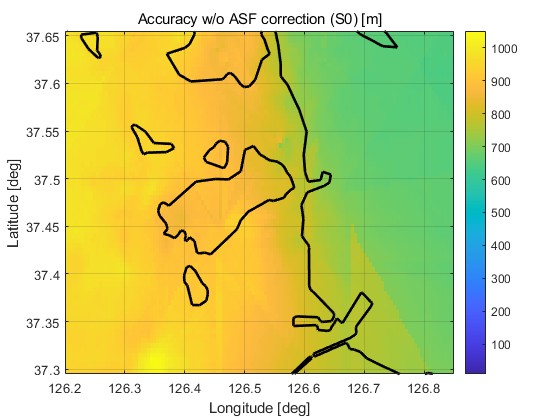}
    \caption{ACC map—S0 (No spatial ASF correction). Spatial ASF offsets remain in the measurements; high-error regions are widely observed.}
    \label{fig:S0}
\end{figure}

\begin{figure}
    \centering
    \includegraphics[width=0.8\linewidth]{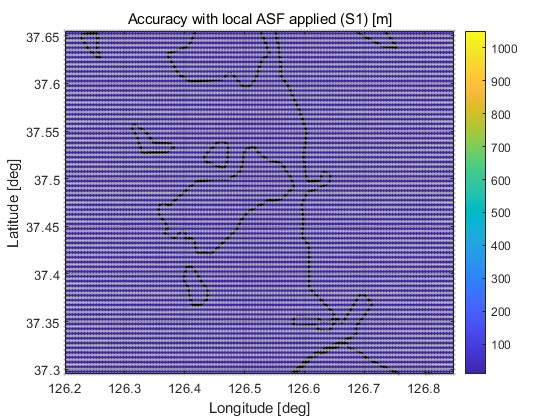}
    \caption{ACC map---S1 (Local spatial ASF application). Measured spatial ASF is removed at each grid point ($\delta\mathrm{ASF}\approx 0$), yielding the lowest errors over the area.}
    \label{fig:S1}
\end{figure}

\begin{figure}
    \centering
    \includegraphics[width=0.8\linewidth]{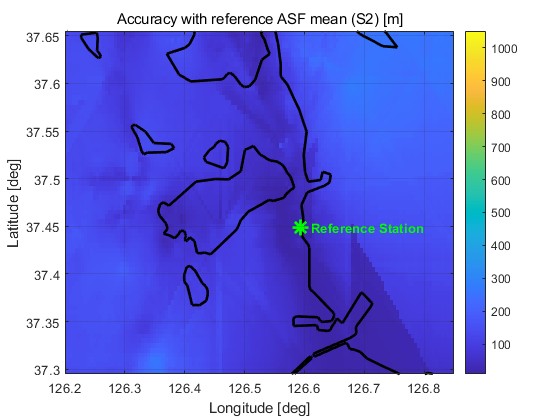}
    \caption{ACC map—S2 (Wide-area single-value broadcast). Reference-station ASF taken from the ASF-map grid nearest to (37.449232°, 126.593994°) is applied network-wide; improvement is concentrated near the reference, with errors increasing away from it.}
    \label{fig:S2}
    
\end{figure}
The comparison of the three ACC maps (S0, S1, S2) yields the following observations.  
Scenario S1 exhibited the lowest mean positioning error across the entire area, showing the most favorable distribution.  
Scenario S0 produced widespread high-error regions and overall the largest errors.  
Scenario S2 demonstrated improvements close to those of S1 in the vicinity of the reference station; however, as distance increased (or along directions with a steep ASF spatial gradient), the errors grew again, and the gap with S0 diminished rapidly.  
The extent and intensity of high-error regions followed the order S0 $>$ S2 $>$ S1. In particular, the error bands and clusters observed in S0 were significantly reduced or disappeared entirely in S1.  

In summary, the maps clearly show that \textbf{local ASF correction (S1)} is the most effective method, whereas \textbf{single-value wide-area correction (S2)} provides improvements only near the reference station.  

\subsection{Comparison of ACC Maps}

When comparing the three ACC maps under the same color scale, the mean positioning errors were:  
S0 = 1154 m, S1 = 10 m, and S2 = 125 m.  
Therefore, S1 reduced the mean error by 1144 m (approximately 99.13\%) compared to S0, and by 115 m (approximately 92\%) compared to S2.  

\section{Conclusion}

This study investigated spatial ASF correction performance for eLoran positioning using measured ASF maps as ground truth. 
Through simulation with three transmitters (Pohang, Gwangju, Socheong), three correction scenarios were compared under identical conditions. 
The results demonstrated that local ASF correction (S1) offers the most effective improvement, reducing the mean positioning error by approximately 60\% compared to no correction (S0) and 39\% compared to wide-area correction (S2). 
While wide-area correction provides partial benefits near the reference station, its accuracy degrades rapidly with distance and ASF spatial variation. 
Therefore, precise spatial ASF mapping is essential for achieving reliable eLoran positioning performance across diverse environments.

\section*{Acknowledgment}

Generative AI (ChatGPT, OpenAI) was used solely to assist with grammar and language improvements during the manuscript preparation process.  
No content, ideas, data, or citations were generated by AI.  
All technical content, methodology, analysis, and conclusions were written and verified solely by the authors.

\bibliographystyle{IEEEtran}
\bibliography{mybibfile, IUS_publications}

\vspace{12pt}

\end{document}